\documentclass[submission,copyright,creativecommons]{eptcs}
\providecommand{\event}{THedu'11} 
\usepackage{epsfig}
\usepackage{amsfonts}

\title{Automatic Deduction in Dynamic Geometry using Sage}
\author{Francisco Botana
\institute{Departamento de Matem\'atica Aplicada I, Universidad de Vigo, Campus A Xunqueira, 36005 Pontevedra, Spain}
\email{fbotana@uvigo.es}
\and
Miguel A. Ab\'anades
\institute{CES Felipe II, Universidad Complutense de Madrid, 28300 Aranjuez, Spain}
\email{abanades@ajz.ucm.es}
}
\def\titlerunning{Automatic Deduction using Sage}
\def\authorrunning{F. Botana and M. A. Ab\'anades}
\begin{document}
\maketitle

\begin{abstract}
We present a symbolic tool that provides robust algebraic methods to handle automatic deduction tasks for a dynamic geometry construction. The main prototype has been developed as two different worksheets for the open source computer algebra system Sage, corresponding to two different ways of coding a geometric construction. In one worksheet, diagrams constructed with the open source dynamic geometry system GeoGebra are accepted. In this worksheet,
Groebner bases are used to either compute the equation of a geometric locus in the case of a locus construction or to determine the truth of a general
geometric statement included in the GeoGebra construction as a boolean variable. In the second worksheet, locus constructions coded using the common file
format for dynamic geometry developed by the Intergeo project are accepted for computation. The prototype and several examples are provided for testing. Moreover, a third Sage worksheet is presented in which a novel algorithm to eliminate extraneous parts in symbolically computed loci has been implemented. The algorithm, based on a recent work on the Groebner cover of parametric systems, identifies degenerate components and extraneous adherence points in loci, both natural byproducts of general polynomial algebraic methods. Detailed examples are discussed.
\end{abstract}

\section{Introduction}

The name Dynamic Geometry Software (DGS) is given to the computer
applications that allow exact on-screen drawing of (generally) planar geometric
diagrams and, their main characteristic, the manipulation of these diagrams
by mouse dragging certain elements making all other elements to automatically
self adjust to the changes. This is also known as interactive geometry.

Since the appearance of the French \textit{Cabri} \cite{cabri} and the American \textit{The Geometer's Sketchpad} \cite{GSP} in the late 80's, many other DGS have been created with slightly different functionalities (C.a.R. Euklides, Dr. Genius, Dr. Geo, Gambol, Geometrix, Geonext,...). Special mention deserve \textit{Cinderella} \cite{cinderella} for its use of complex numbers methods as basis for its computations and \textit{GeoGebra} \cite{GeoGebra}, whose open source model and effective community development has resulted in a spectacular world wide distribution.

From the beginning, DGS have been the paradigm of new technologies applied to Math education. Being able to produce a great number of examples of a configuration has many times been taken as a substitute for a formal proof in what has come to be known as a \emph{visual proof}. Questions have been raised on the influence of this use of DGS on the development of the concept of proof in school curricula \cite{proofGDS}. This is a symptom of the incompleteness of general DGS relative to further manipulation of configurations. Although most DGS considered come equipped with some property checker, their numeric nature does not really provide a sound substitute for a formal proof.

To compensate the computational limitations of DGS, two main approaches have been taken to add symbolic capabilities to DGS. Some systems incorporate their own code to perform symbolic computations (e.g. \cite{GEX}), while other systems, including several by the authors, choose to reuse existing Computer Algebra Systems (CAS) (e.g. \cite{LAD, BotanaValcarceCAE, RoanesBridge, Geother}). The prototype presented here offers a solution in this latter direction with a relevant change: it is strictly based on open source tools.

Open source development, with its philosophy resembling a bazaar of different agendas and approaches, was thought at the beginning to be only good for small applications. Its success, clearly exemplified by the magnitude of the operating system GNU/Linux, came as a surprise to even the most optimistic programmers, who thought that big applications would always need a reverent cathedral-building approach. Nowadays, open source applications have impacted all computing areas and are no longer considered marginal.

If we take open as a synonym for accessible, we can not find fields where this concept is more relevant than Education, where universal access should be the leading principle, and Mathematics, where public scrutiny is at its very foundations. In consequence, the open source nature of the Math educative prototype that we present here, is not only a characteristic but a statement.

To finish this section, a few words on the main elements of our system, namely GeoGebra, Sage and Intergeo, are provided as a basic introduction. 

GeoGebra is an open source DGS with algebraic capabilities, establishing a direct relationship between the objects in the different windows: graphics, algebra, and spreadsheet. While its technical characteristics are of first order, what really made it our DGS of choice in this project was its impressive community integration that makes GeoGebra a \textit{de facto} standard in the field.

GeoGebra was created in 2001 by Markus Hohenwarter at the University of Salzburg (Austria). It won the 2002 Academy Award European software (EASA) in the category of Mathematics and since then, using the word of mouth and Internet, its use spread rapidly throughout the world. It has become a collaborative project with impressive figures: users in 190 countries (The UN has 192 member states!), versions in 48 languages and half a million monthly visitors to its web site.

Sage is an open source Computer Algebra System designed to be a viable free open source alternative to proprietary systems such as Maple, Mathematica or Matlab. The integration of multiple tools, the possibility of remote access via the Internet and the emphasis for decency and freedom make its most notable features. Sage was created in 2004 by William Stein \cite{stein}, professor at the University of Washington, after several disagreements over some accessibility issues with the developers of Magma, a highly specialized commercial CAS in whose development he had collaborated.

Sage is built out of nearly 100 open-source packages (including Singular and Maxima). It has a unified user interface that takes the form of a notebook in a web browser or the command line. Using the notebook, you can connect either locally to your own Sage installation or remotely to a Sage server. One remarkable characteristic of the use of Sage over a server is the possibility to share worksheets with other users. For its power and versatility we foresee Sage as the de facto standard for teaching mathematics with computers in secondary and university levels.

The \textit{Intergeo} (i2g) file format is a specification based on the markup language XML designed to describe constructions created with a DGS. It is one of the main results of the Intergeo project, an eContentplus European project dedicated to the sharing of interactive geometry constructions across boundaries. For more information about the project, we refer to http://i2geo.net and the documentation available there, as well as to \cite{intergeo_documentation_1, intergeo_documentation_2}.

An intergeo file takes the form of a compress file package. The main file is intergeo.xml, which provides a textual description of the construction in two parts, the elements part describing a (static) initial instance of the configuration and the constraints part where the geometric relationships are expressed. For more details on the file format the reader is referred to \cite{intergeo_format}, where several examples are provided.

Among the list of elements covered by the i2g format, we find several definitions of locus, used in our prototype. However no analogue of the GeoGebra boolean statement is considered (yet) by the i2g format. This is the reason why the Sage worksheet that computes the equation of geometric loci specified with the i2g format does not accept true/false queries.

Details on how to use and test the system and several examples are provided in sections \ref{Prototype_Description} and \ref{sect:Examples}. Finally, a special version of the environment for GeoGebra is presented in section \ref{sect:Automatically_detecting}. It incorporates a new protocol by the authors to eliminate extraneous parts in symbolically computed loci. The algorithm, based on a recent work on the Groebner cover of parametric systems, identifies degenerate components and extraneous adherence points in loci, both natural byproducts of general polynomial algebraic methods. 

\section{Automatic deduction and locus equations}\label{sect:Automatic_deduction_and_locus_equations}

\begin{figure}[ht!]
\centerline{\fbox{\psfig{figure=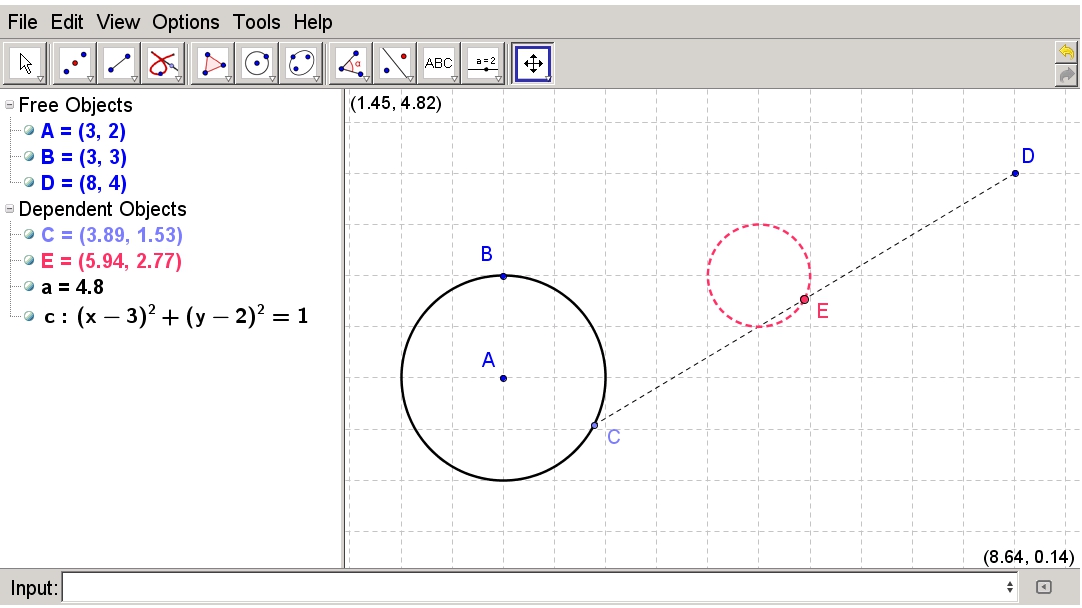,width=8cm}}}
\caption{Locus determined by the midpoint of $CD$ as $C$ runs along the circle.}
\label{fig:locus_example}
\end{figure}

A distinctive characteristic of dynamic geometry environments is the possibility of obtaining the plot of all the points determined by the different positions of a point, the tracer, as a second point in which the tracer depends on, called the mover, runs along the one dimensional object to which it is restrained. This set of points is called a locus. Figure \ref{fig:locus_example} shows the locus determined by the midpoint of the segment $CD$ as the mover $C$ runs along the circle with center $A$.

For most DGS a locus is basically a set of points in the screen with no algebraic information. This prevents the system from working with the locus set as a standard element in the construction. Even when the locus is a simple curve (such as the circle in figure \ref{fig:locus_example}) for which certain geometric commands are defined (such as tangent-line), the fact that the locus set is not really viewed as an algebraic curve by the system makes these operations undefined. 

Two main approaches can be considered to determine the implicit equation of a locus set. The numerical approach, based on interpolation, has been followed, for instance, by the commercial systems Cinderella and Cabri. However, its numerical nature makes it subject to inaccuracies (see \cite{BotanaICCS2002} for details).

On the other hand there is the symbolic method followed in this work. In a dynamic geometry construction with algebraic elements (i.e. defined by polynomials), a locus is determined by the solution of a parametric polynomial system where the parameters correspond to the symbolic coordinates of the mover point. Finding the equation of a locus can be viewed then as \textit{discovering} the algebraic properties (i.e. polynomial equations) that the (coordinates of the) tracer point must satisfy in order for the construction to keep its geometric constraints while the mover point changes its position along its path. In other words, the computation of a locus (of its equation that is) can be viewed as a particular instance of automatic discovery in geometry.

The algorithm for automatic discovery of loci followed in this work was first proposed in \cite{BotanaValcarceMatCom2003} and it derives from an earlier proposal for automatic discovery in elementary geometry via algorithmic commutative algebra and algebraic geometry using Groebner bases in \cite{RecioVelez99}. This same algorithm has recently been implemented by the DGS JSXGraph to determine the equation of a locus set using remote computations on a server (\cite{JSXGraphADG2010}), an idea previously developed by the authors (\cite{LAD}). 

Roughly speaking, the procedure for automatic discovery goes as follows. A statement is considered where the conclusion does not follow from the hypotheses. Symbolic coordinates are assigned to the points of the construction (where every free point gets two new free variables $u_i, u_{i+1}$, and every bounded point gets up to two new dependent variables $x_j, x_{j+1}$) so the hypotheses and thesis are rewritten as polynomials $h_1,\dots,h_n$ and $t$ in $\mathbb{Q}[u,x]$. Eliminating the dependent variables in the ideal $(hypotheses, thesis)$, the vanishing of every element in the elimination ideal $(hypotheses, thesis)\cap \mathbb{Q}[u]$ is a necessary condition for the statement to hold.

The problem of finding the equation of a locus for a particular geometric construction can be viewed as a particular case of this general setting with numerical coordinates for the free points, free variables corresponding to the tracer point and dependent variables to all other bounded points. The elimination (using Groebner bases) of the dependent variables in the polynomial ideal obtained as translation of the construction leaves us with a set of polynomials in the independent variables. The zero set corresponding to these polynomials correspond to a superset of the sought locus set. This algebraic set may contain extra components due sometimes to the algebraic nature of the method (it returns only Zariski closed sets) and due sometimes to some degenerate positions in the construction. A solution for these special situations recently developed by the authors is discussed in section \ref{sect:Automatically_detecting}.

\section{Prototype Description}\label{Prototype_Description}

\begin{figure}[ht!]
\centerline{\fbox{\psfig{figure=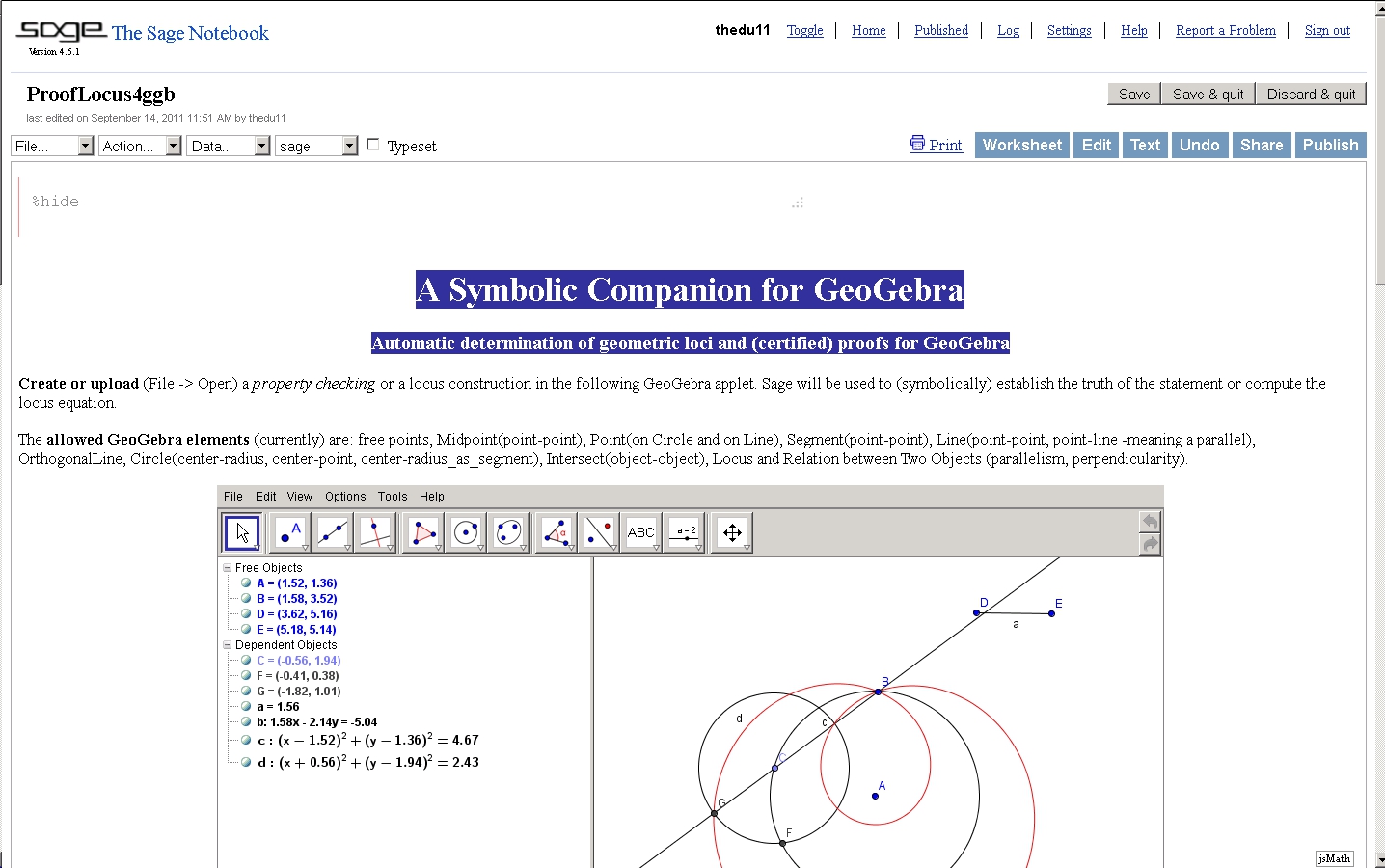,width=8cm}}}
\caption{Sage worksheet with GeoGebra applet.}
\label{fig:SageInterface}
\end{figure}

The main element of the automatic deduction environment prototype consists of a Sage worksheet (LocusProof4ggb.sws) in which two different tasks are performed over GeoGebra constructions: the computation of the equation of a geometric locus in the case of a locus construction, and the study of the truth of a general geometric statement included in the GeoGebra construction as a Boolean variable. Both tasks are implemented using algebraic automatic deduction techniques based on Groebner bases computations. The direct communication Sage-GeoGebra is made possible via the inclusion of a GeoGebra applet in the Sage worksheet inside  HTML code. This applet allows the direct construction of a geometric diagram or the upload of a previously designed GeoGebra construction to the Sage worksheet (see figure \ref{fig:SageInterface}).

More precisely, once a geometric diagram has been input in the applet, a JavaScript method provided by GeoGebra is used to obtain the GeoGebra XML description of the construction as a string inside a JavaScript variable. However, Sage does not include any standard way to transfer the content of a JavaScript variable to a Sage variable. The solution to this problem comes from \textit{tampering} with the code of Sage itself, using the following code (inside some HTML element) to control the flow of the different Sage cells. \vspace{2 mm}

{\small
\begin{verbatim}
    code = "myxml = " + " \'\'\'" + document.applets[0].getXML() + " \'\'\' "; 
    \$("#cell_input_17").val(code); 
    evaluate_cell(17);
\end{verbatim}
}

\vspace{2 mm}

This makes the XML description of the GeoGebra construction available to Sage as a regular string. Then a first parsing process takes place in which the different GeoGebra elements in the construction are read, creating an algebraic counterpart for each element. This is done with some Python code involving several hundred lines of code. As an example, the following snippet correspond to the definition of the \textit{PerpendicularLine} function that generates the equation of the perpendicular line to a given line through a given point. \vspace{2 mm}

\begin{center}
{\small
\begin{verbatim}
    def PerpendicularLine(n,p,l):
        """Constructs the line $n$ perpendicular to the line $l$ through $p$."""
        vdir=Todo[l]['vdir']
        eq=Set([(absc - x(p))*vdir[0] + (orde - y(p))*vdir[1]])
        vdir=(-vdir[1],vdir[0])
\end{verbatim}
} 
\end{center}

\vspace{2 mm}

Once all the algebraic structures in the construction have been set up, the appropriate variables are initialized and the ideal corresponding to the task (locus or proof) is generated. Singular (a CAS included in Sage with special emphasis on commutative algebra) is then called to basically compute a Groebner basis for this ideal. Each generator is factored and a process of logical expansion is performed on the conjunction of the generators in order to remove repeated factors. 

The way the answer is presented to the user depends on the task. In the case of a locus, besides providing the locus equation, the graph corresponding to this equation is included in the GeoGebra applet together with the original locus construction. In the case of a proof, a logical statement about the truth of the statement is provided. More details together with some remarks on the information provided by the different answers can be found in the description of the examples in section \ref{sect:Examples}.

The system is presented in a prototype state and currently accepts a limited (easy to expand) set of construction primitives, namely Free points, Midpoint (point-point), Point (on Circle and on Line), Segment (point-point), Line (point-point, point-parallelLine), OrthogonalLine, Circle(center-radius, center-point, center-radiusAsSegment), Intersect (object-object), Locus and RelationBetweenTwoObjects (parallelism, perpendicularity).

Something worth remarking is the fact that the system is solely based on Open Source software, in contrast with other implementations of DGS-CAS communication based on proprietary applications (e.g. \cite{BotanaValcarceCAE} with Mathematica, \cite{RoanesBridge} with Maple, \cite{GeometryExpressions} with Maple and/or Mathematica).

The main Sage worksheet described above depends on a particular DGS. To make the algorithms available to other systems a variation of the prototype has been developed that admits locus constructions specified using the more general intergeo common file format. In this case, no graphic interface is provided and the (intergeo) XML-description of the locus povided by the DGS has to be directly included in a text area. 

We find several definitions of locus in the list of elements covered by the i2g format. However, no analogue of a GeoGebra boolean statement is considered by the format. This is the reason why the developed Sage worksheet does not accept true/false queries. 

In \cite{intergeoImplementationTable} one can find the list of intergeo compliant DGS and their current compatibility status. An example using the DGS JSXGraph (\url{http://jsxgraph.de}) has been included in section \ref{sect:Examples}.

Both worksheets are available for testing online in the main Sage server (\url{http://www.sagenb.org/}, user: THedu11, password: test). They also have been made publicly available from the authors account in that same server (\cite{LocusProof4ggbThedu11} and \cite{Locus4i2gThedu11}).

It has to be noted that the communication from GeoGebra to Sage is not synchronous. That is, changing an element in the GeoGebra construction does not automatically trigger the corresponding update in the CAS side. This technical improvement will be accomplished as further work.
\section{Examples}\label{sect:Examples}

We illustrate the use of the prototype with three examples: the proof of a basic theorem in the form of a GeoGebra construction with a boolean variable and the computation of two loci specified in GeoGebra and with the i2g coding. These and other examples are available at \url{http://webs.uvigo.es/fbotana/THedu11/} together with some demo videos.

\subsection{A GeoGebra example of proof}

\begin{figure}[ht!]
\centerline{\fbox{\psfig{figure=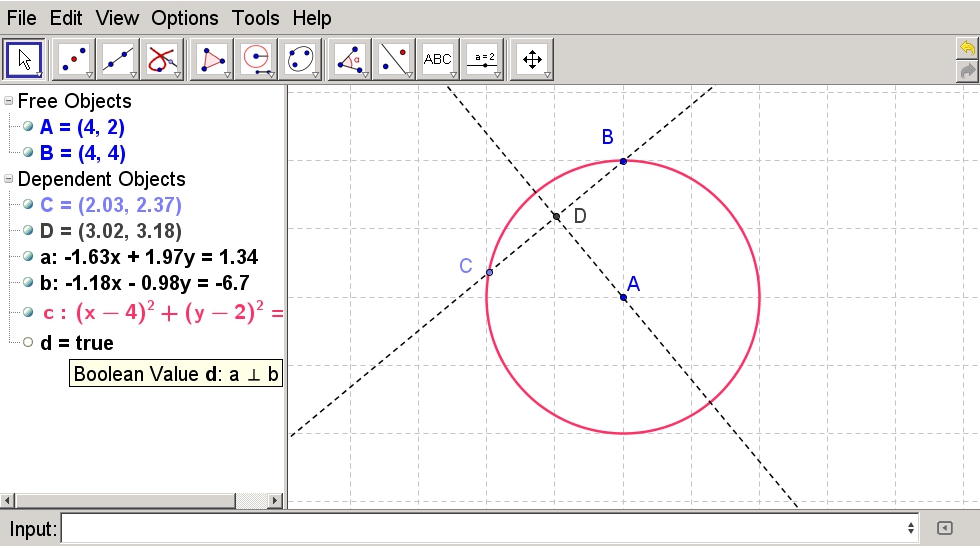,width=8cm}}}
\caption{The radius through the midpoint of a chord is perpendicular to the chord.}
\label{proof_circle_ggb}
\end{figure}

We consider the basic theorem that states that given a circle and a circle chord, the radius through the midpoint of the chord is perpendicular to the chord. This result can be easily reproduced in terms of a boolean variable in a GeoGebra construction. Given the circle with center $A$ through $B$, it suffices to consider a point $C$ in the circle, the midpoint $D$ of the segment $BC$ and the line joining $A$ and $D$. The boolean statement encompassing the theorem is $line(A,D) \bot line(B,C)$ (see figure \ref{proof_circle_ggb}). 

\begin{figure}[ht!]
\centerline{\fbox{\psfig{figure=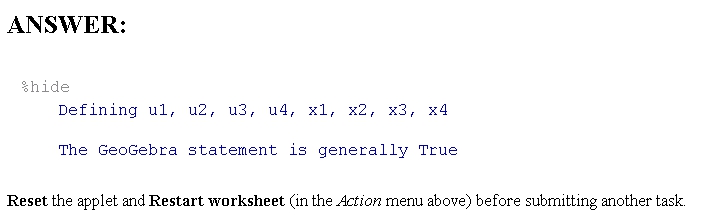,width=9cm}}}
\caption{The radius through the midpoint of a chord is perpendicular to the chord.}
\label{proof_circle_sage}
\end{figure}

The \emph{true} provided by GeoGebra based on numerical computations is symbolically corroborated by Sage as shown in figure \ref{proof_circle_sage}.

Notice that, unlike the answer provided by GeoGebra, the answer provided by Sage is not only based on symbolic computations but also completely general.
Symbolic variables ($u_1$ to $u_4$) are used as generic coordinates for the center $A$ and the point $B$ defining the circle, what makes the answer a general statement about any chord of any generic circle. The sentence in the answer includes the technical expression \textit{generically true} meaning that the statement is true when there is a sensible construction, excluding situations such as that in which points $A$ and $B$ coincide, for instance.    

\subsection{A GeoGebra example of locus}

We consider the construction of a simple locus determined by a moving point on a circle. Starting with the circle with center $A$ through $B$ and a point $C$ on the circle, we consider the locus of the points $D$ intersection of the line perpendicular to $AC$ through $B$ and the line parallel to $AB$ through $C$. 

\begin{figure}[ht!]
\centerline{\fbox{\psfig{figure=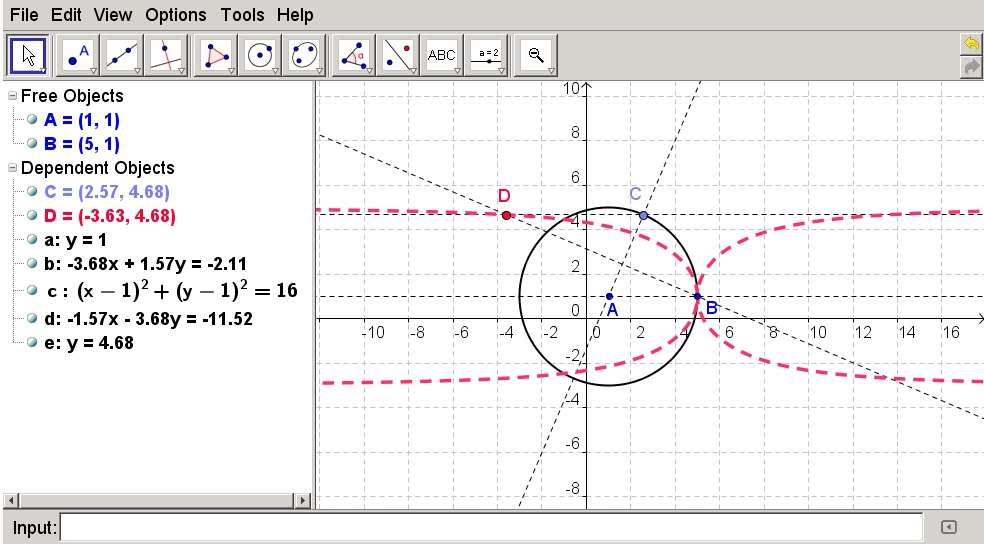,width=8cm}}}
\caption{Locus determined by $D$ as $C$ runs along the circle.}
\label{fig:2LikeParabolas_ggb}
\end{figure}

In figure \ref{fig:2LikeParabolas_ggb} we can see the two branches of the locus as plotted by GeoGebra. Note that the left branch of the locus corresponds to $C$ being in the right-hand-side part of the circle while the right branch corresponds to $C$ being in the left-hand-side part of the circle. 

\begin{figure}[ht!]
\centerline{\fbox{\psfig{figure=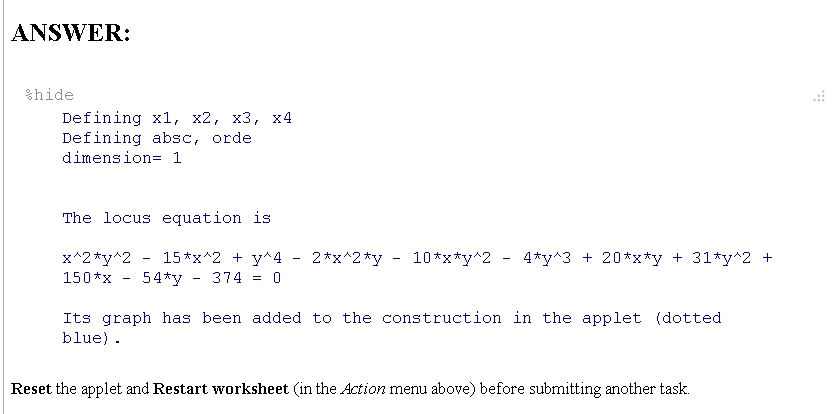,width=9cm}}}
\caption{Equation of locus as provided by Sage.}
\label{fig:2LikeParabolas_equation}
\end{figure}

To the eyes of a non experienced user (such as a secondary education student), the branches of the locus could seem like two parabolas, a guess that is not possible to check in GeoGebra. Figure \ref{fig:2LikeParabolas_equation} shows the exact equation for the locus set provided by Sage establishing the non parabolic nature of the locus.

\begin{figure}[ht!]
\centerline{\fbox{\psfig{figure=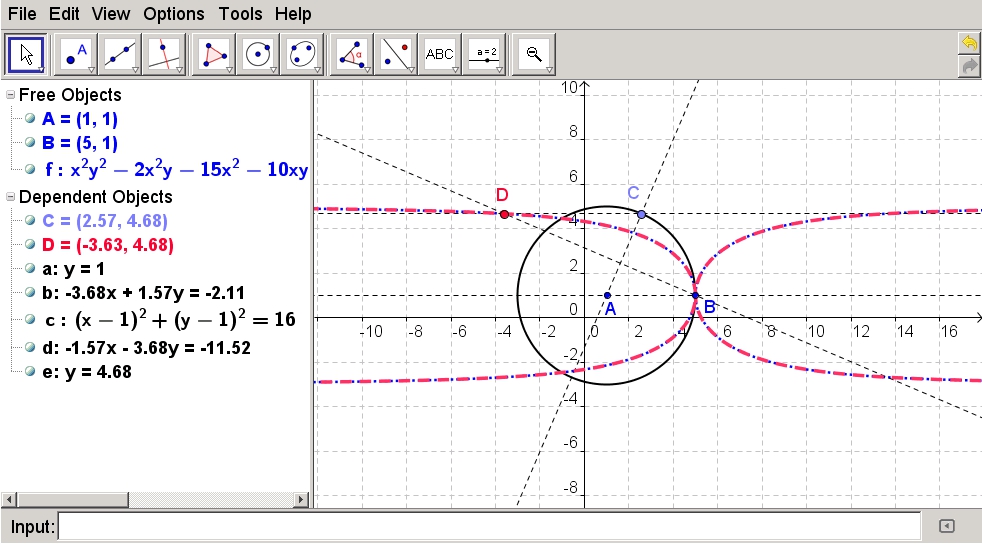,width=8cm}}}
\caption{Locus determined by $D$ as $C$ runs along the circle.}
\label{fig:2LikeParabolas_answer_applet}
\end{figure}

The implicit curve corresponding to this equation is automatically included in the GeoGebra applet together with the original representation of the locus set confirming the correctness of the computations (see figure \ref{fig:2LikeParabolas_answer_applet}).  

\subsection{An Intergeo example of locus}
To obtain a locus description, one can use any of the DGS supporting the i2g format (see \cite{intergeoImplementationTable} for a list of softwares and their current implementation status). We consider the example provided in the Intergeo web site as an example of a construction with the \verb"locus_defined_by_point_on_circle" element. It is the construction, in JSXGraph, of a cardioid as the locus set traced by a point $P$ as a point $X$ runs along a circle \cite{cardioid}. In fact, the i2g description provided by JSXGraph had a double definition for the intersection point $X$ that had to be changed eliminating its definition as a free point. 

As a sample, the following is the encoding of the locus element in the i2g description of the construction where one can read the names of the locus set, the tracing and the moving points as well as some technical information to be interpreted by the DGS. 

\begin{verbatim}
<locus_defined_by_point_on_circle>
  <locus out="true">L</locus>
  <point>X</point>
  <point>P</point>
  <circle>c</circle>
</locus_defined_by_point_on_circle>
\end{verbatim}

\begin{figure}[ht!]
\centerline{\fbox{\psfig{figure=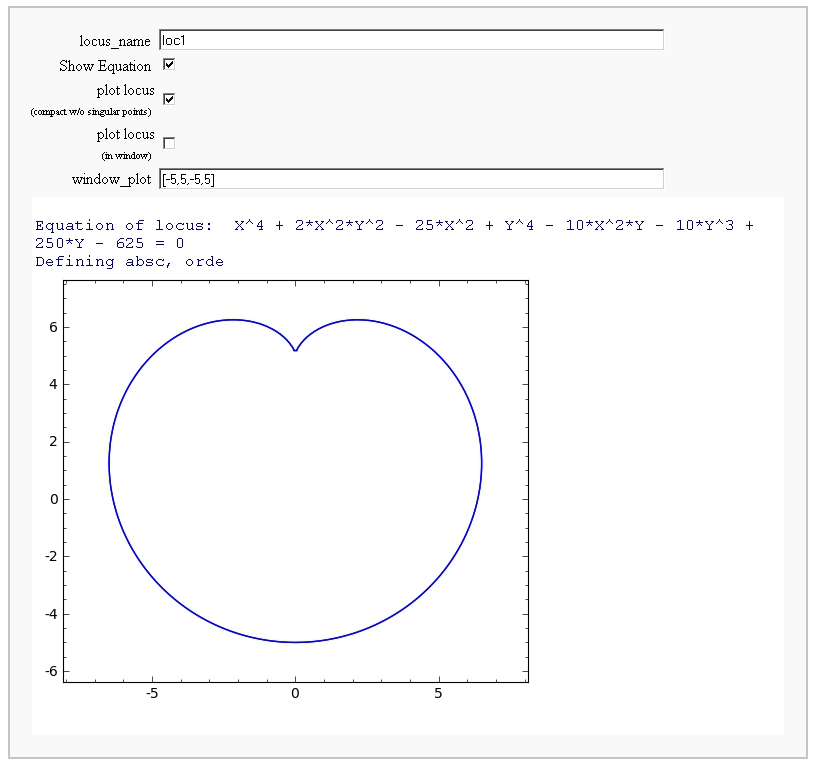,height=7.0cm}}}
\caption{Equation and graph of cardioid as given by Sage.}
\label{locus_i2g}
\end{figure}

Figure \ref{locus_i2g} shows the equation and graph of the locus provided by Sage.

\section{Automatically detecting degenerate components of loci}\label{sect:Automatically_detecting}

As recalled in section \ref{sect:Automatic_deduction_and_locus_equations} a locus set is, for most DGS, a set of points on the screen. This numerical approximation to the generation of loci has limitations such as the lack of adequate heuristics for an efficient interpolation of sample points and the impossibility of knowing a reliable algebraic description. The latter makes, for instance, impossible the computation of tangents to loci, a standard command in other curves. 

As shown by the examples in section \ref{sect:Examples}, the symbolic approach solves these problems providing a polynomial characterization of a locus. However, this approach introduces two new problems. On the one hand, the translation to polynomials of the geometry of the locus can introduce extra algebraic constraints unrelated to the original geometric construction. On the other hand, extraneous solutions can also be introduced due to the elimination procedure. 

Both issues (removing degenerated parts and returning loci as constructible sets, rather than varieties) can be efficiently addressed in the field of dynamic geometry by using the theory of parametric polynomial systems. The variables occurring in the equations which describe a construction can be naturally divided into a set of parameters and a set of unknowns. The parameters correspond to the coordinate of the free and bounded points, while the unknowns (variables), correspond to the coordinates of the locus point. Thus, the parametric polynomial system is formed by the polynomial geometric constraints. We seek then solutions of the parametric system in terms of the parameter values and are interested in the structure of the solution space. 

In \cite{BotanaAbanades2011}, the authors have developed an algorithm based on the recent work on the Groebner cover of parametric systems presented in \cite{MontesWibmer2010} that efficiently identifies degenerate components and extraneous adherence points in computationally computed loci.

We omit the details of the main method and just mention that the GrobnerCover algorithm is used in such a way that the roles commonly assigned to parameters and variables is exchanged, allowing the removal of degenerate components and spurious parts in the adherence.

In this section we present a prototype of a Sage-GeoGebra deduction environment similar to the one illustrated in section \ref{sect:Examples} in which this algorithm has been implemented. No further technical details will be given about the Locus$\_$nondeg$\_$thedu11.sws Sage worksheet (available for testing online in \url{http://www.sagenb.org/}, user: THedu11, password: test, and also at \cite{LocusNondeg4ggbThedu11}). Two examples of use will be shown instead. 

\subsection{Example 1: removing extra components}\label{subsect:removing_extra_components}

\begin{figure}[ht!]
\centerline{\fbox{\psfig{figure=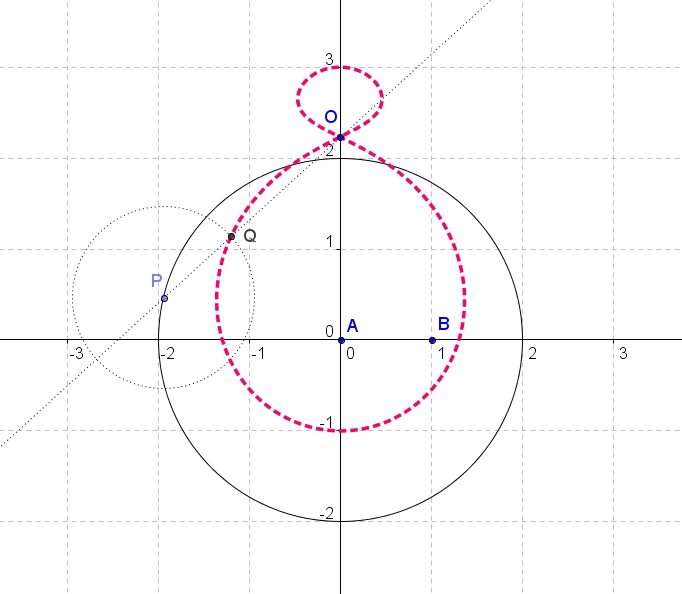,width=5.0cm}}}
\caption{Locus set traced by $Q$ as $P$ runs along its circle.}
\label{fig:Limacon_punto_fuera_ggb}
\end{figure}

Let us consider the conchoid constructed as a locus set as follows. Consider a point $O$ in the plane, a circle $c$ and the line $l$ passing through $O$ and $P$ (any point on c). Let $Q$ be a point on $l$ such that $distance(P,Q)=k$, where $k$ is a constant. The conchoid is the locus set traced by $Q$ as $P$ moves along $c$ as shown in figure \ref{fig:Limacon_punto_fuera_ggb} (where $k$ is given by the length of segment $AB$, note the circle around $P$ determined by $k$).  

\begin{figure}[ht!]
\centerline{\fbox{\psfig{figure=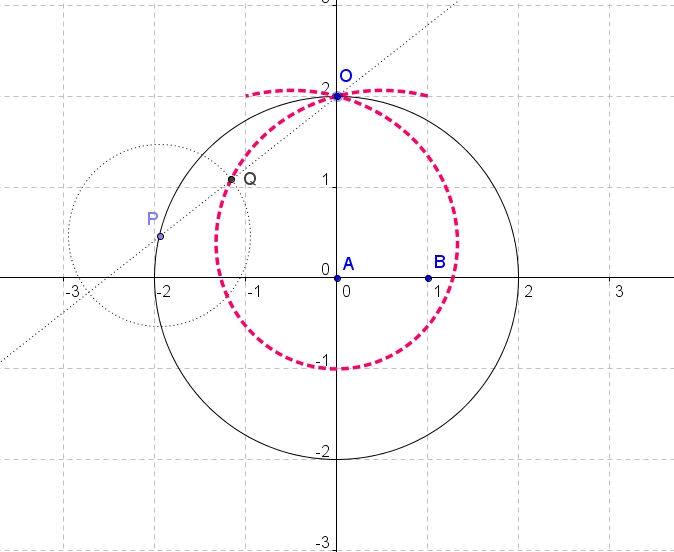,width=5.0cm}}}
\caption{Lima\c{c}on of Pascal as the locus set traced by $Q$ as $P$ runs along the circle $c$.}
\label{fig:Limacon_punto_dentro_ggb}
\end{figure}

In this construction, as $P$ moves along the circle, all elements are well defined and no degenerate case is encountered. But if we consider a variation of this locus set by making the point $O$ a point in the circle $c$, the situation changes dramatically. We obtain the locus set known as the lima\c{c}on of Pascal (named after Blaise Pascal's father) partially shown in figure \ref{fig:Limacon_punto_dentro_ggb}.

Note that to obtain the complete graph of the lima\c{c}on of Pascal in GeoGebra a second locus would have had to be defined considering as tracer the other intersection point between line $OQ$ and the auxiliary circle centered at $P$.

\begin{figure}[ht!]
\centerline{\fbox{\psfig{figure=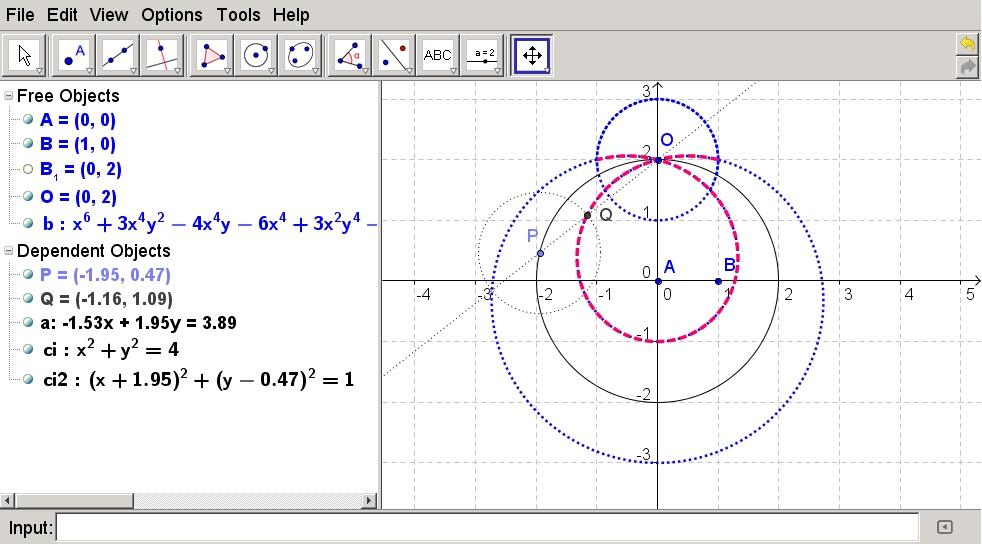,width=8.0cm}}}
\caption{Lima\c{c}on of Pascal with extra circle.}
\label{fig:Limacon_con_extra}
\end{figure}

\begin{figure}[ht!]
\centerline{\fbox{\psfig{figure=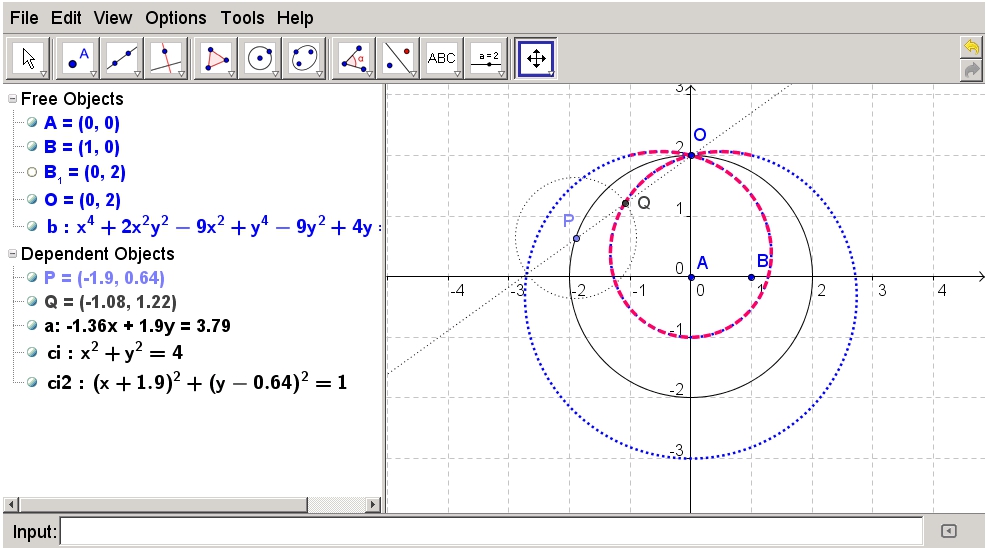,width=8.0cm}}}
\caption{Lima\c{c}on of Pascal without the (automatically removed) extra circle.}
\label{fig:Limacon_sin_extra}
\end{figure}

To symbolically compute the equation of this set, the following assignment of coordinates is made: $P(a,b)$, $Q(x,y)$. If moreover we consider that $O$ is the point $(0,2)$ and $distance(A,B)=1$ we get, as polynomial description of the construction, the ideal $I=\langle a^2+b^2-4,(x-a)^2+(y-b)^2-1, x(b-2)-a(y-2) \rangle $ whose generators correspond, respectively, to the following geometric constraints: $P$ is in the circle of center $(0,0)$ and radius $2$, $distance(P,Q)=1$ and $Q\in Line(P,O)$. Eliminating the variables $a$ and $b$, a product of two polynomials is obtained: $(x^4 + 2x^2y^2 + y^4 - 9x^2 - 9y^2 + 4y + 12)(x^2 + y^2 - 4y + 3)$. The first factor provides the implicit equation for the actual lima\c{c}on while the second factor corresponds to a spurious circle corresponding to the degenerate case for which $P=O$ when the line $l$ ceases to exist. Figure \ref{fig:Limacon_con_extra} shows the (complete) lima\c{c}on with the extra circle as provided by the Sage worksheet LocusProof4ggb$\_$thedu11.sws while figure \ref{fig:Limacon_sin_extra} shows the final result provided by the worksheet Locus$\_$nondeg$\_$thedu11.sws after automatically detecting the extra component and removing it.

\subsection{Example 2: removing extra adherence points}\label{subsect:removing_extra_adherence_points}

\begin{figure}[ht!]
\centerline{\fbox{\psfig{figure=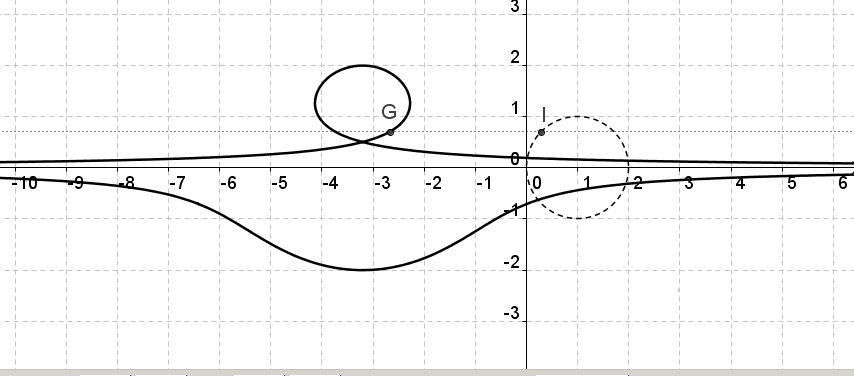,width=8.0cm}}}
\caption{Moving point $G$ on Nichomedes conchoid and its horizontal projection $J$ onto circle.}
\label{figure:locus_nichomedes_projection_ggb}
\end{figure}

Consider the moving point $G$ on the Nichomedes conchoid. The horizontal line through $G$ defines the locus set of points $I$ on this line and on the circle of radius $1$ centered at $(1,0)$ as shown in figure \ref{figure:locus_nichomedes_projection_ggb}).

The standard elimination procedure states that the locus set is the whole circle $(x-1)^2+y^2=1$, whereas the true locus is the projection of the conchoid on this circle, that is, the circle without the origin and the point $(2,0)$. Being $I$ the ideal generated by the construction polynomials, $V=\mathbf{V}(I)$ the variety generated by $I$, $I_{x,y}$ the ideal that eliminates all variables but those of the locus point, and $\pi_{x,y}(V)$ the sought projection, we have that $\pi_{x,y}(V) \subseteq \mathbf{V}(I_{x,y})$. So, through elimination we get the smallest variety containing the locus, or, in others words, the Zariski closure of the projection. 

\begin{figure}[ht!]
\centerline{\fbox{\psfig{figure=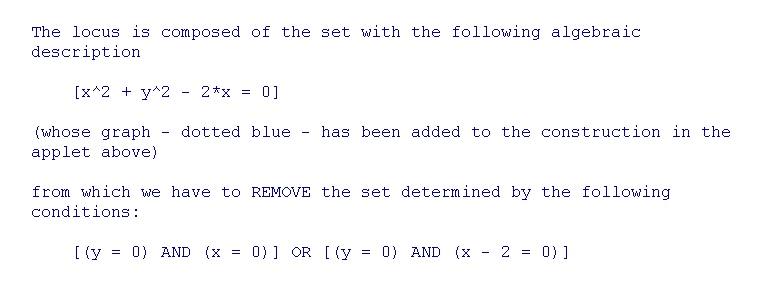,width=10.0cm}}}
\caption{Exact description of locus set as a constructible set as provided by Sage.}
\label{figure:locus_nichomedes_projection_sage}
\end{figure}

However, for the same example, in Locus$\_$nondeg$\_$thedu11.sws we obtain the description of the locus set as a constructible set, namely $\mathbf{V}(\langle (x-1)^2+y^2-1 \rangle) \backslash \mathbf{V}(\langle x(x-2),y\rangle)$. This information is included in the answer provided by Sage as shown in figure \ref{figure:locus_nichomedes_projection_sage} where a textual notation of set unions and intersections has been followed to describe the set to be removed.

\section{Conclusions}\label{sect:Conclusions}

The interactive environment joining a dynamic geometry system and a computer algebra system presented in this article follows a simple idea also found in other authors. However, this presentation presents two main innovations. First, at the technical level, the system consists of an efficient complete integration of two different open source applications via a bidirectional communication between the applications. Second, at the theoretical level, a system to automatically remove unwanted elements in an automatically computed locus is presented. The system tackles two subtle problems inherent to the nature of algebraic polynomial methods in DGS, namely the identification of degenerate components and extra adherence points in a locus. 

We find important to remark that the fact that the system is solely based on readily available open source tools make it completely accessible to a general user.

\subsection*{Acknowledgements}
The authors thank Rado Kirov for his helpful indications on Sage programming and acknowledge the financial support by research
grant MTM2008-04699-C03-03/MTM from the Spanish MICINN.


\bibliographystyle{eptcs}
\bibliography{AutomaticDeduction}

\end{document}